\documentclass[useAMS,usenatbib]{mn2e}
\usepackage{epsfig}
\usepackage{psfig}

\title[A survey for low luminosity quasars at redshift $z\sim5$] {A
  survey for low luminosity quasars at redshift $z\sim5$}
  \author[Sharp R.G., Crampton D., Hook I.M. and McMahon R.G.] {Sharp
  R.G.$^1$,\thanks{E-mail : rgs@ast.cam.ac.uk}\thanks{Visiting
  Astronomer, Canada-France-Hawaii Telescope operated by the National
  Research Council of Canada, the Centre National de la Recherche
  Scientifique de France and the University of Hawaii.}, Crampton
  D.$^2$\footnotemark[2], Hook I.M.$^3$ and McMahon
  R.G.$^1$\footnotemark[2].\\ $^1$Institute of Astronomy, Madingley
  Road, Cambridge, CB3 0HA, UK\\ $^2$Dominion Astrophysical
  Observatory, HIA, National Research Council of Canada, Victoria,
  B.C. V9E 2E7, Canada\\ $^3$Department of Physics, University of
  Oxford, Nuclear \& Astrophysics Laboratory, Keble Road, Oxford, Ox1
  3RH, UK\\ e-mail:rgs@ast.cam.ac.uk}

\pagerange{\pageref{firstpage}--\pageref{lastpage}} \pubyear{2002}

\def\LaTeX{L\kern-.36em\raise.3ex\hbox{a}\kern-.15em
    T\kern-.1667em\lower.7ex\hbox{E}\kern-.125emX}

\def\PaA{\ifmmode \mathrm{Pa}{\alpha}\else Pa$\alpha$\fi}
\def\PaB{\ifmmode \mathrm{Pa}{\beta}\else Pa$\beta$\fi}
\def\PaG{\ifmmode \mathrm{Pa}{\gamma}\else Pa$\gamma$\fi}

\def\Lya{\ifmmode \mathrm{Ly}{\alpha}\else Ly$\alpha$\fi}
\def\Nv{N\,\textsc{v}}
\def\etal{\emph{et al.\thinspace}}
\def\sqdeg{deg$^{2}$}

\begin{document}

\label{firstpage}

\maketitle

\begin{abstract}
We present the results of a multi-colour ($VIZ$) survey for
low luminosity (M$_B$$<-23.5$) quasars with $z\sim5$ using the
12K CCD mosaic camera on CFHT. The survey covers 1.8\sqdeg \ to a limiting
magnitude of m$_z=22.5$(Vega), about two magnitudes fainter than the 
SDSS quasar survey. 20 candidates were selected by their $VIZ$ colours and 
spectra for 15 of these were obtained with GMOS on the Gemini North telescope. 
A single quasar with $z=4.99$ was recovered, the remaining candidates are 
all M stars.

The detection of only a single quasar in the redshift range accessible
to the survey (4.8$<$5.2) is indicative of a possible turn over in the
luminosity function at faint quasar magnitudes, and a departure from
the form observed at higher luminosities (in agreement with quasar
lensing observations by Richards \etal (2003)).  However, the derived
space densitys, of quasars more luminous than M$_{B}$(Vega)$<$-23.5, of
2.96$\times$10$^{-7}$ Mpc$^{-3}$ is consistent at the 65\% confidence
level with extrapolation of the quasar luminosity function as derived
by Fan \etal (2001a) at m$_i$$<$19.6(Vega).
\end{abstract}


\section{Introduction}
High redshift quasars provide direct probes of conditions near the epoch of
re-ionization when galaxies were first being assembled. Thus far only the most
luminous members of the high redshift population of quasars have been studied,
most notably with the Sloan Digital Sky Survey (SDSS York \etal (2000)). 
However, little is known about the space density of quasars at
lower luminosities

At high luminosities (M$_B<-25$) Fan \etal (2001a) use observations from
the  SDSS over the redshift
range $4<z<5$ to define a quasar luminosity function.  Space density
estimates based on the detections of the first quasars at $z>5.8$ (Fan
\etal (2003)) are incompatible with the strong evolution model proposed
for the quasar luminosity function by Warren, Hewett and Osmer
(1994). A primary goal of the present survey is to probe more than two magnitudes
fainter than SDSS.

Two of the major uncertainties in the role of quasars in the early
Universe are the transition luminosity between the observed steep
bright end of the luminosity function and the slope of the lower
luminosity region of the distribution function, which is constrained
to be shallower to prevent the over production of intermediate mass
relic black holes at lower redshift.

Both the Big Throughput Camera 40\sqdeg\ survey (BTC40) of Monier
\etal (2002) and the Wide Field Survey of the Issac Newton Telescope
(INTWFS) (Sharp, McMahon, Irwin and Hodgkin (2001), Sharp (2002))
probe the high redshift ($z>4.6$) quasar luminosity function to
greater depths than that reached by SDSS.  These surveys cover
sufficient area to recover five quasars with $z>4.6$ to date.  Both
surveys report results consistent with extrapolation of the single
power law approximation to the luminosity function derived by Fan
\etal (2001a) with little evidence for any turn over to M$_{B}<-25$.
If such a turn over exists, and assuming a double power law form for
the luminosity function (as presented by Boyle \etal (2000) for the
2df quasar survey at $z<3$), then it can be shown that for a non
divergent faint end slope of the luminosity function (L$^\beta$,
$\beta>$-2)) the major contribution of quasars to the ionising UV
background at high redshifts is emitted at the knee point of the
function, an analogue to the fiducial parameter of the galaxy
luminosity function, L$^*$.

In this work we present the results of a new survey program to probe
the quasar luminosity function to M$_B$$<$-23.0 at $z\sim5$ with a
view to constraining the faint end slope of the quasar luminosity
function below L$^*$.

Conventional Vega magnitudes are used throughout this work.  Except
where explicitly stated to the contrary
a cosmology is adopted throughout such that H$\rm_0$=70 Km s$^{-1}$
Mpc$^{-1}$, and $\Omega_{\rm{M}}$=0.3, $\Omega_{\rm\Lambda}$=0.7.

\section{Observations : Optical Imaging data}
The imaging data were obtained with the 3.6m Canada-France-Hawaii
(CFH) Telescope and the CFH12K camera.
CFH12K is a close packed mosaic of 12 back side illuminated
MIT Lincoln Laboratories CCDs giving a total field of view in a single
observation of 42$\times$28arcmin and a pixel size of 0.206arcsec.
The cosmetic quality of the mosaic is good with only one chip
significantly affected by bad columns.  Coupled with the excellent
seeing routinely obtained at CFHT, the fine pixel scale allows
excellent morphological classification of objects leading to a
reduction of the number of marginally extended non quasar sources in
the candidate list.

\begin{figure}
 \psfig{file=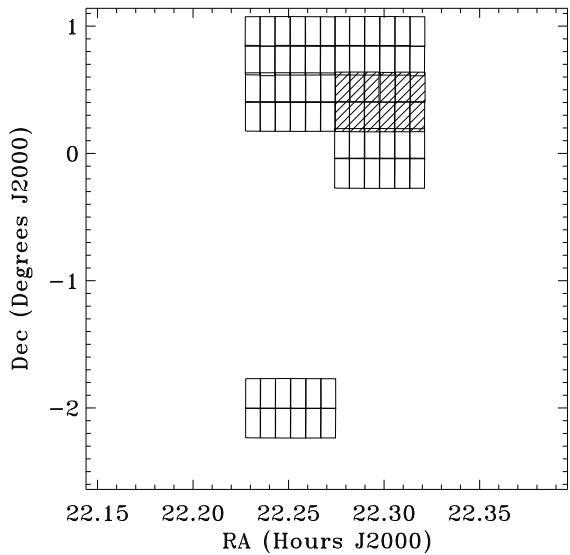,width=8cm}
 \caption{\label{cfdf field} The location of individual CCD footprints
 for the 6 pointing observation of the CFDF 22hr field
 are shown.  The total survey area, observed in $VI$ and $Z$ bands, is
 1.77\sqdeg.  Observational details are given in table \ref{obsdata}
 and filter response curves are shown in Figure \ref{cfht viz
 filters}.  The shaded region represents the 0.3\sqdeg \ field with
 deeper observations.}
\end{figure}

Observations were obtained in the $VI$ and $Z$ bands during the six
nights 12-17 September 1999 interleaved with observations for
alternate projects.  Details are given in table \ref{obsdata}.  Figure
\ref{cfdf field} shows the pointing layout for the Canada-France Deep
Field (CFDF: McCracken \etal 2001)) 2215+00 field from which the
observations presented here are drawn.  Anderson \etal (2001)
subsequently identified a high luminosity $z=4.99$ quasar within the
2215+00 field using data from SDSS.  The recovery of this objects as a
candiadte quasar is discussed in section \ref{pSDSSJ2216+0013, a
quasar at z=4.99}.

The data form part of an extended survey project of $\sim$6.2\sqdeg \
observed in the $VI$ and $Z$ bands.  $I$ band imaging from the public
NOAO survey\footnote{http://www.archive.noao.edu/ndwfs/} field at
02:07$-$04:30(J2000) will be used in conjunction with $V$ and $Z$ band
imaging obtained with CFH12K.  Candidate selection and spectroscopic
followup observations have been carried out on a 1.8\sqdeg \ region of
the survey centred on the Canada-France Deep field region at
22:45$+$00:30(J2000) for which $VI$ and $Z$ data have been obtained.

\begin{figure}
 \psfig{file=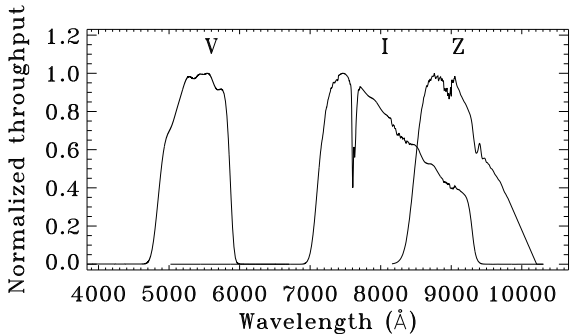,width=8cm}
 \caption[Filter pass bands]{\label{cfht viz filters}Filter pass bands
 used for the observations.  Throughputs have been normalised to unity
 for comparison.  Models of atmospheric transmission and CCD quantum
 efficiency have been included with the filter response functions.}
\end{figure}

Figure \ref{cfht viz filters} shows the filter transmission curves for
the filters used in the survey.  With accurate star/galaxy separation,
possible here due to the excellent seeing during the observations and
the well matched pixel scale of the CFH12K camera, the dominate
contaminant population within candidate lists for most high redshift
quasar colour selection surveys is low mass stars.  The extended red
tail of the $I$ filter is therefore not ideal for this work since the
overlap between the $I$ and $Z$ filters reduces the power of the $I-Z$
colour index to identify low mass stars.  However, the $VIZ$ filter
set, and the associated colour-colour diagram (Figure \ref{cfht viz}),
does provide sufficent discrimination between quasars at redshift
$z>4.8$ and low mass stars.

\begin{table*}
\caption[]{\label{obsdata}Observational details.}
\begin{tabular}{lcclrrr}
\textbf{Field} &
\textbf{RA Dec (J2000)} &
\textbf{Area} &
\textbf{Observations} &
\multicolumn{3}{c}{\textbf{Depth 5$\bf{\sigma}$ (Vega)$\dagger$}} \\
\hline
\hline
Canada-France Deep Field & 22:17 $+$00:24 & 0.31\sqdeg  & $V\times$4h, $I\times$1h, $Z\times$1h    & m$_V<$26.1 & m$_I<$24.5 & m$_Z<$23.3\\
\hline
Canada-France Deep Field & 22:17 $+$00:24 & 1.21\sqdeg & $V\times$1h, $I\times$0.5h, $Z\times$0.5h & m$_V<$25.5 & m$_I<$24.0 & m$_Z<$23.1\\
Anon-22                  & 22:15 $-$02:00 & 0.31\sqdeg & $V\times$1h, $I\times$0.5h, $Z\times$0.5h & m$_V<$25.5 & m$_I<$24.0 & m$_Z<$23.1\\
\hline
NOAO Deep field          & 02:07 $-$04:30 & 4.25\sqdeg & $V\times$1h, $Z\times$0.5h                & m$_V<$25.5 &            & m$_Z<$23.1\\
NOAO Deep Wide-Field Survey & 02:10 $-$04:30& 4.25\sqdeg & ---                                     &            & m$_{I}<$25.5$\dagger$\\              
\hline
\end{tabular}
\raggedright Notes: $\dagger$ Depths are 5$\sigma$ derived in
0.8arcsec radius (median seeing) aperture.  For the 02:07$-$04:30 NOAO
field, $I$ band data from the public NOAO Deep Wide-Field Survey will
be used to complement CFH12K $V$ and $Z$ band observations.  The $I$
band NOAO limit is 5$\sigma$ in 2arcsec aperture taken from
http://www.noao.edu/noao/noaodeep/.
\end{table*}

\subsection{Imaging data reduction}
The data were processed using \textsc{IRAF} and the \textsc{MSCRED}
package.  Twilight flat field frames are required to facilitate the
removal of interference fringes from the $I$ and $Z$ data.  Fringe
removal is carried out by iteratively scaling and subtracting a master
fringe frame to minimise residual background variations.  The process
is automated through the use of software created as part of the Isaac
Newton Telescope Wide Field Camera (INTWFC) data reduction pipeline
(Irwin and Lewis (2001)).  The master fringe frame is compiled on a
nightly basis from flat fielded sky observations.  Fringing in the $Z$
band is present at the 7-10\% level before processing.  Residual
fringing remains at the 1-2\% level prior to image alignment and
stacking. Figure 3 gives an example of the results of the data
reduction procedure.

Object catalogue generation, morphological classification and the
computation of an astrometric solution was performed using elements of
the INTWFC pipeline (Irwin and Lewis 2001)) and with the aid of the
\textsc{GAIA} image analysis
tool\footnote{http://star-www.dur.ac.uk/$\sim$pdraper/gaia/gaia.html}.
Aperture photometry is derived based on a median seeing radius
aperture (0.8arcsec) and is used for $VI$ and $Z$ images.  An aperture
correction is derived on a chip-by-chip bases from a curve of growth
analysis of bright, non saturated stars in the field.  The World
Coordinate System (WCS) is used to merge the object catalogues across
the $VI$ and $Z$ filters with the $Z$ catalogue used as the reference.
Flux calibrations is achieved with observations of standard star
fields from the Landolt (1992) catalogue (primarily L95 and L110).
Observations were recorded at several locations across the CCD mosaic.

\begin{figure}
\centering
 \psfig{file=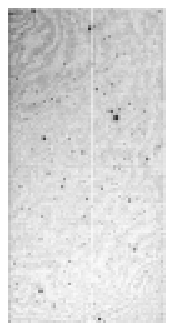,width=1.6cm}
 \psfig{file=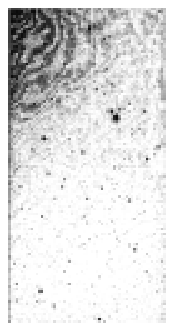,width=1.6cm}
 \psfig{file=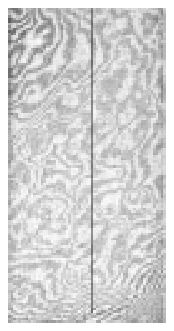,width=1.6cm}
 \psfig{file=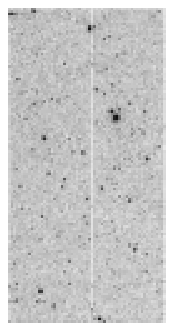,width=1.6cm}
 \psfig{file=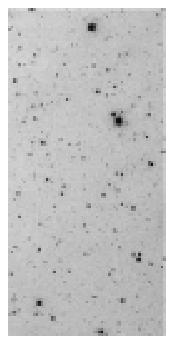,width=1.6cm}
 \caption[CFH12K fringe removal (chip7)]{\label{Chip 7 fringe} Example
 of the defringing process applied to $I$ and $Z$ band imaging data.
 A sequence of $Z$ band frames are shown.  All images are scaled to a
 common gray-scale level with the exception of the left-most frame
 which is shown with an extended scale.  Images are (left to right);
 Flat fielded frame with fringing, frame 1 repeated with gray-scale
 common to remaining images, master fringe frame, defringed frame and
 finally the stacked frame (combining 6$\times$600sec exposures of the
 CFDF region).}
\end{figure}

\section{Colour selection of candidates}
\label{colour selection}
At $z>3$ the optical colour indices of quasars are dominated by the
unabsorbed non-thermal blue continuum longward of \Lya\
($\lambda$1216\AA) in the quasar rest frame and the effects of neutral
hydrogen absorption in the Inter Galactic Medium (IGM) shortward of
the \Lya\ emission line.  At redshift $z\sim5$ the spectral break
across \Lya\ falls between the $V$ and $I$ bands.

Figure \ref{cfht viz} shows a colour-colour diagram derived from
CFH12K imaging in the $VIZ$ bands for a 0.2\sqdeg\ subsection of the
CFDF field.  The predicted quasar colour track, as a function of
redshift, is over plotted.  The colour track is computed assuming an
underlying quasar spectrum based on a power law with spectral index
$S_\nu\propto \nu^{-0.5}$ and with an emission line spectrum based on
the composite spectrum of Vanden Berk \etal (2001).  The absorption
model for the intervening \Lya\ forest (IGM) is taken from
Madau(1995).  The stellar main sequence is clearly visible as the
heavily populated strip in the centre of the plot.  All objects
classified as stellar and with m$_{Z}<$22.5 are shown.

Candidate high redshift objects are identified in the diagram as
stellar objects lying below the stellar locus.  To establish the
selection criteria the stellar locus in the colour-colour diagram is
approximated by a linear fits of the form,

\begin{equation}
(V-I) = m \times (I-Z) + c,\\
\end{equation}

an offset from the locus is then chosen so as to minimise the stellar
contamination of the candidate list while maximising the accessible
redshift range.  The colours selection criterion then becomes,

\begin{equation}
\begin{array}[1]{l}
	(V-I) > 0.22, (I-Z) < -0.1,\\
	(V-I) = 6.06 \times (I-Z) + 0.83, (I-Z) < 0.6,\\
\end{array}
\end{equation}

Once selected as a candidate quasar, candidates are visually inspected
to check the validity of the photometry.  This step is required to
identify spurious sources such as objects in the diffraction spikes of
bright stars.  Two example candidates are shown in Figure \ref{sdss
quasar}.  A finding chart is also generated based on the $Z$ band
image at a larger scale.  Image defects such as stellar halos and poor
fringe correction are readily identified on inspection of these
charts.  Such candidates are flagged as lower priority accordingly.
While the image quality of the CFH12K mosaic is generally high, this
step greatly reduces the number of candidates requiring spectroscopic
observation.  Since targets are selected as outliers in colours space,
we are particularly vulnerable to objects with spurious photometry.

\subsection{Stellar classification requirements}
Candidate high redshift quasars are expected to appear as unresolved
(PSF limited) point sources in $I$ and $Z$ band observations with
little or no detectable flux in the $V$ band due to absorption in the
IGM.

Candidate selection based on classification as a point source
in the $Z$ band is
required to reduce the target list to a practical size for
observation.  Clearly a $V$ band classification cannot be used since
classification becomes unreliable at the survey limit and is
impossible for drop out objects. Classification based on the $I$ band
would be preferable due to the reduced residual fringing when compared
to the $Z$ band.  However, a problem with the focus mechanism of the
camera during observations resulted in a degradation of the image
quality (a slight ellipticity) for $I$ band observations.  Object
classification is therefore based on $Z$ band imaging with $I$ band
classification used only as a secondary indicator for assessing
quasar candidates.The actual classification of an object as a point source 
is derived from a curve of growth analysis from
object fluxes recorded in progressively larger radii apertures centred
on each source (Irwin and Lewis (2001)).

The assumption of a stellar classification for quasars requires that
there will be no significant contamination from the quasar host
galaxy.  Even at the lower redshift cutoff for $VIZ$ colour selection
($z>4.8$) the $Z$ band filter samples the rest frame UV spectrum
short-ward of 2000\AA\ and any such contamination would be negligible.

\subsection{Dust extinction}
Any quasar selection based on rest frame UV colour indicies is
subject to bias against objects suffering moderate dust extinction
either along the line of sight (Fall and Pei (1993)) or intrinsic to
the quasar (Sharp \etal (2002)).  In order to draw a comparison with
the luminosity function of Fan \etal (2001a) no attempt is made to
account for possible selection effects from dust reddening
in the current work.

\begin{figure}
 \psfig{file=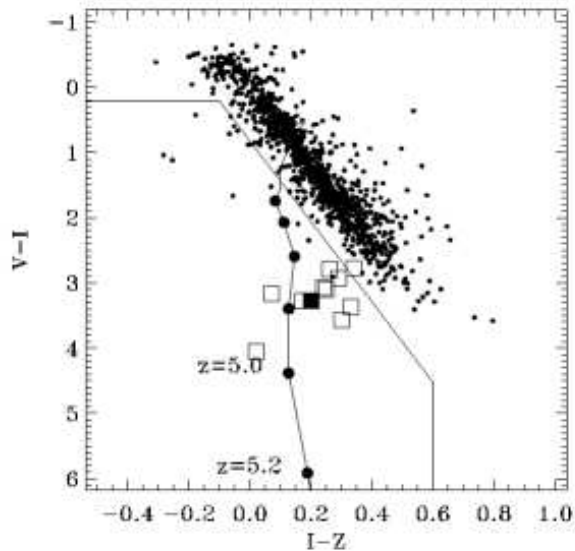,width=8cm}
 \caption[CFH12K VIZ candidates selection]{\label{cfht viz} Candidate
 selection is performed based on the colour-colour diagram constructed
 from the $VI$ \& $Z$ bands.  The theoretical quasar locus is marked
 at redshifts z=5.0 and 5.2 (filled circles indicate lower redshifts
 with $\Delta$z=0.2). Open symbols mark candidate quasars drawn from
 the full 1.8\sqdeg\ survey. The filled square represents the known
 z=4.99 quasar pSDSSJ2216+0013.  The stellar locus is compiled from a
 0.2\sqdeg \ subsection ($\sim$10\%) of the CFDF survey area.}
\end{figure}

\begin{figure}
\centering
 \psfig{file=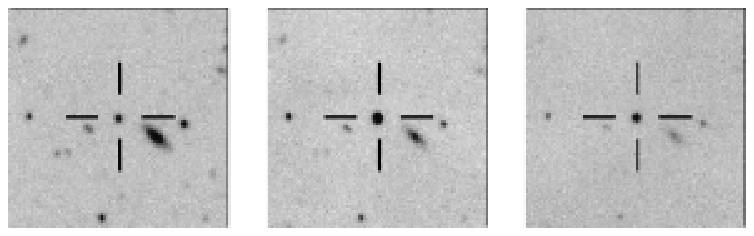,width=9cm}\\
\centering
\psfig{file=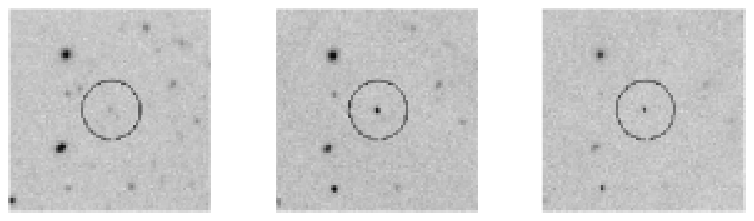,width=8cm}
\caption[Known SDSS high redshift quasar]{\label{sdss quasar} $VI$ and
$Z$ band images are shown from left to right.  The field size in each
image is 40arcsec.  Similar image sequences of all candidate quasars
are examined to check the validity of photometry.\\
Upper : The rediscovered SDSS high redshift quasar pSDSSJ2216+0013 at
z=4.99.\\
Lower : Identified as a quasar candidate by it's very red,
$V-I=3.57$, colour index, cfh12k8 (Table \ref{candidates}) is
identified spectroscopically as an early type M star.}
\end{figure}

\section{pSDSSJ2216+0013, a quasar at z=4.99}
\label{pSDSSJ2216+0013, a quasar at z=4.99} A previously known high
redshift quasar was rediscovered in the field.  Quasar pSDSSJ2216+0013
at z=4.99 (Anderson \etal (2001)) is readily identified as a candidate
quasar in the $VIZ$ colour-colour diagram (Figure \ref{cfht viz}).
The CFDF was not preselected for observation based on the presence of
pSDSSJ2216+0013 within the field.

Figure \ref{sdss quasar} shows the $VIZ$ images for this object.  The
observational details are given in table \ref{sdss quasar table} and
are compared with the reported observations of pSDSSJ2216+0013 from
the SDSS.  No new spectroscopic observations of pSDSSJ2216+0013 have
been performed.

The spectrum of pSDSSJ2216+0013 is unremarkable, with the exception of
the high redshift and bright magnitude.  The discovery spectrum
(Anderson \etal\ (2001)) shows the typically strong continuum break
across a strong narrow \Lya/\Nv \ blend supporting the photometric
measurements.  Given these sharp features and the slight mismatch
between the SDSS and CFH12K filter systems, plus potential for quasar
variability, the observed magnitudes and colours are in good
agreement.  The coordinate offset between the reported SDSS position
and that measured from CFH12K is $\sim$0.28arcsec (-0.2arcsec in each
axis).  This is at the predicted limit of the external accuracy of the
survey WCS computed with the INTWFC data processing pipeline.  This
accuracy is primarily limited by the accuracy with which unsaturated
objects within the imaging data can be tied to a well defined WCS from
astrometric survey data.  Internally to the survey the astrometric
accuracy between different filter observations is at the
$\sim$0.1arcsec level.

\begin{table}
\centering
 \caption[Data on pSDSSJ2216+0013 at z=4.99]{\label{sdss quasar table}The
 rediscovered SDSS high redshift quasar pSDSSJ2216+0013.}
\begin{tabular}{|lcc|}
\hline
\textbf{Name}            & \multicolumn{2}{l}{pSDSSJ2216+0013}\\
\textbf{Redshift}	& \multicolumn{2}{l}{4.99}\\
\textbf{M$\bf \rm_{B}$}	& \multicolumn{2}{l}{-26.6}\\
\hline
\hline
\textbf{SDSS} & \textbf{sinh\,AB} & \textbf{Vega}\\
r 		& 21.78 & 21.60\\
i 		& 20.30 & 19.88\\
z 		& 20.33 & 19.77\\
i-z		& -0.03 & 0.11\\
Ra Dec (J2000)  & \multicolumn{2}{c}{22:16:44.02 +00:13:48.3}\\
\hline
\textbf{CFH12K} & & \textbf{Vega}\\
V 		& --- & 23.01\\
I 		& --- & 19.73\\
Z 		& --- & 19.56\\
V-I		& --- & 3.28\\
I-Z		& --- & 0.20\\
Ra Dec (J2000)  & \multicolumn{2}{c}{22:16:44.034 +00:13:48.51}\\
\hline
\end{tabular}
\end{table}

\section{Observations : spectroscopic}
\label{Observations : spectroscopic}
Observations of fifteen quasar candidates were obtained with the GMOS
spectrograph (Murowinski \etal (2003) in preparation, Hook \etal
(2003) submitted) during regular queue-scheduled observations on
the Gemini North telescope in November 2001 and June-July 2002. The
November observations were among the first regular science
observations with GMOS North and, as a result, could not be taken with
the proper blocking filter to remove the second order spectra. Conditions were
photometric or near-photometric for the majority of the observations
and the image quality was better than 1 arcsec. A 0.75\arcsec slit
was used with the R400 grating and the detector was binned 2$\times$2,
yielding a dispersion of 1.4 \AA/pixel and a spectral resolution
(determined by the slit width) of 9\AA. The useful spectral range was
typically 5900-9500\AA. An OG515 filter was used to remove the second
order for the 2002 observations. A single exposure was obtained for
most objects, with exposure times ranging from 1800s to 2400s (for the
faintest targets).

The data were reduced following standard methods with the Gemini IRAF
tools\footnote{http://www.gemini.edu/sciops/instruments/gmos/gmosIndex.html}
to pre-process and mosaic the spectra from the 3 CCD detectors
together, followed by extraction and reduction of the 1D spectra with
\textsc{IRAF SPECRED}. An approximate flux calibration was determined
from observations of standard stars with identical instrument
parameters during the same period of observation (contamination from
second order blue light further reduced the accuracy of the flux
calibration for the 2001 November observations).

\subsection{Observed spectra of candidate quasars}
Details of the fifteen candidate quasars are given in table
\ref{candidates}.  No strong emission line objects are detected.
Indeed, all of the observed spectra show strong similarities and are
suggestive of early type M stars.  In order to determine the
classification of stars prone to selection using the $VIZ$ colour
diagram a composite stellar spectrum is constructed by co-adding the
individual observations of each candidate (Figure \ref{all spectra
combined}).

The composite spectrum is examined following the classification
schemes of Kirkpatrick, Henry and McCarthy (1991) (for K-M stars) and
Kirkpatrick \etal (1999) (for later type L stars).  The presence of
pronounced absorption features, coincident with the expected location
of TiO bands, suggest identification with stars later than $\sim$M2.
The lack of a broad potassium feature at $\sim$7700\AA\ is taken as
evidence that the spectra are not from L stars. Absorption in VO bands
becomes increasingly pronounced in later M stars.  We therefore
identify the composite spectra with a classification earlier than M5.

Examination of the spectra of each target individually supports the
classification range (M2-5) assigned to the composite spectra.  TiO
absorption bands are readily identified.  While the VO feature is in a
region of the spectrum affected by strong OH sky residuals, there is
little evidence for significant absorption.

A classification for stellar interlopers in the range M2-M5 is
consistent with the synthetic photometry prediction shown in Figures
\ref{low mass stars iz} and those of Dobbie, Pinfield, Jameson and
Hodgkin (2002).

\begin{table*}
\raggedright
\caption[] {\label{candidates}Details of the fifteen candidate
spectroscopically observed. The primary target list contains twenty
such targets, suggesting a 75\% follow up completeness.}
\centering
\begin{tabular}{lcrrrrr}
\textbf{Object} &
\textbf{RA Dec (J2000)} &
\multicolumn{3}{c}{\textbf{Magnitude (Vega)}} &
\multicolumn{2}{c}{\textbf{Colour (Vega)}}\\
 & & \textbf{V} & \textbf{I}& \textbf{Z} & \textbf{V-I} & \textbf{I-Z}\\
\hline
\hline
cfh12k1  & 22:15:55.38, $+$00:58:25.4 & 25.01 & 22.23 & 21.89 & 2.78 & 0.34\\        
cfh12k2  & 22:18:30.50, $+$00:59:59.3 & 25.39 & 22.28 & 22.03 & 3.10 & 0.25\\	     
cfh12k3  & 22:18:06.14, $+$00:42:11.2 & 25.05 & 22.13 & 21.90 & 2.92 & 0.23\\	     
cfh12k4  & 22:18:33.37, $+$00:39:23.7 & $>$25.36 & 22.57 & 22.31 & $>$2.79 & 0.26\\  
cfh12k5  & 22:18:36.88, $+$00:39:30.7 & $>$25.37 & 22.44 & 22.15 & $>$2.93 & 0.29\\  
cfh12k6  & 22:17:39.05, $+$00:33:45.3 & 26.32 & 22.34 & 21.90 & 3.98 & 0.44\\	     
cfh12k7  & 22:16:28.76, $+$00:10:57.0 & 25.24 & 21.97 & 21.80 & 3.26 & 0.17\\	     
cfh12k8  & 22:17:29.54, $+$00:11:46.5 & 25.48 & 21.91 & 21.60 & 3.57 & 0.34\\	     
cfh12k9  & 22:17:47.34, $+$00:15:05.1 & 26.17 & 23.00 & 22.93 & 3.17 & 0.07\\        
cfh12k10 & 22:18:40.03, $+$00:11:55.1 & 24.73 & 21.91 & 21.80 & 2.82 & 0.11\\	     
cfh12k11 & 22:13:40.93, $-$01:49:06.6 & 25.23 & 22.63 & 22.57 & 2.60 & 0.06\\	     
cfh12k12 & 22:15:42.26, $-$02:12:58.3 & 25.49 & 22.69 & 22.49 & $>$2.80 & 0.20\\     
cfh12k13 & 22:18:30.16, $+$00:09:16.9 & 25.56 & 22.20 & 21.87 & 3.37 & 0.33\\	     
cfh12k14 & 22:16:49.87, $-$00:13:44.0 & 25.49 & 22.85 & 22.74 & $>$2.64 & 0.11\\     
cfh12k15 & 22:16:53.65, $-$00:13:33.3 & 25.23 & 22.15 & 21.91 & 3.07 & 0.24\\	     
\hline
\end{tabular}
\end{table*}

\begin{figure}
\centering
\psfig{file=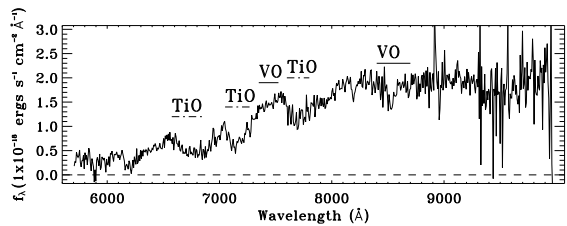,width=7.5cm}
\caption[] {\label{all spectra combined} To provide a preliminary
classification, a simple summation is performed of all the candidate
spectra.  The resulting spectrum is then smoothed and re-binned by a
factor of five (corresponding to the spectroscopic resolution
element).  Following the classification scheme of Kirkpatrick, Henry
and McCarthy (1991) the composite spectrum suggests, as expected from
synthetic $VIZ$ band photometry models, the stellar interlopers fall
in the range of spectral types M2-M5.  TiO absorption bands and the
absence of a strong absorption features due to K suggest
classification as an M star.  The weakness of VO absorption at longer
wavelengths argues against late type M stars.}
\end{figure}

\begin{figure}
\centering
 \psfig{file=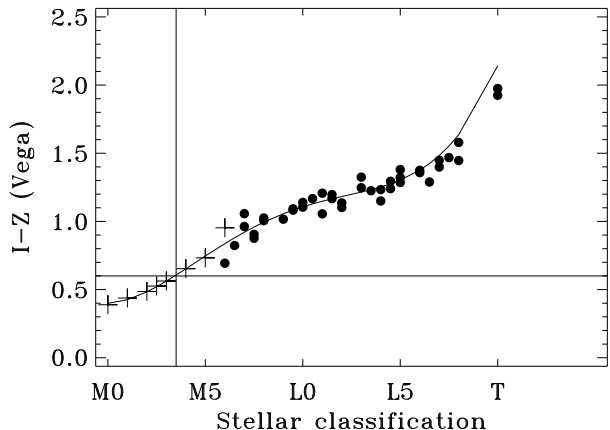,width=8.5cm}
 \raggedright \caption{\label{low mass stars iz} The colour of low
 mass stars as a function of classification can be computed using an
 atlas of stellar spectra.  The $I-Z$ colours, in the CFH12K filter
 system, are in agreement with those found by Dobbie \etal (2002).
 The colours are estimated using the Keck LRIS spectra of Reid \etal
 (2001, http://dept.physics.upenn.edu/$\sim$inr/ultracool.html) for
 stars later than M5 and atlas of Pickles (1998,
 http://archive.ast.cam.ac.uk/viz-bin/VizieR) for M0V-M6V .  Based on
 synthetic photometry measurements, the sample $I-Z<0.6$ colour cut
 selects against stars later than M3.5.}
\end{figure}

\section{Quasar space density}
A tentative quasar space density of 2.96$\times$10$^{-7}$ Mpc$^{-3}$
(65\% confidence limits $\pm_{7.5\times10^{-9}}^{1.6\times10^{-6}}$)
is inferred for the redshift range $4.8<z<5.2$ and absolute magnitude,
M$_B<-23.5$(Vega) limit accessible to the 1.8\sqdeg \ CFH12K $VIZ$
survey.

The survey area is estimated via monte-carlo integration.  The survey
completeness function is estimated in the manner adopted by Sharp
\etal (2001).  A grid of model quasar spectra is generated, over the
accessible redshift range, with parameters for the quasar spectral
model, discussed in section \ref{colour selection}, drawn at random
from the distribution functions
\begin{equation}
\begin{array}[1]{l}
\alpha=N(-0.5,{0.3^2}),\\
EW(\Lya)=N(92.91,\rm{34.0^2})
\end{array}
\end{equation}
Synthetic photometry estimates are computed for each model quasar and
colours are determined within the survey filter system.
Representative survey photometric errors are then added to the
photometry, scaling each spectrum over the magnitude range, m$_z$,
available.  The fraction of quasars recovered by the survey colour
selection criteria, as a function of redshift and magnitude m$_z$, is
recorded as an estimate of the survey completeness function.

To m$_{Z}<$22.5 and in the redshift range 4.8$<$z$<$5.2 the survey is
predicted to be 70\% complete. \\

\begin{figure}
 \psfig{file=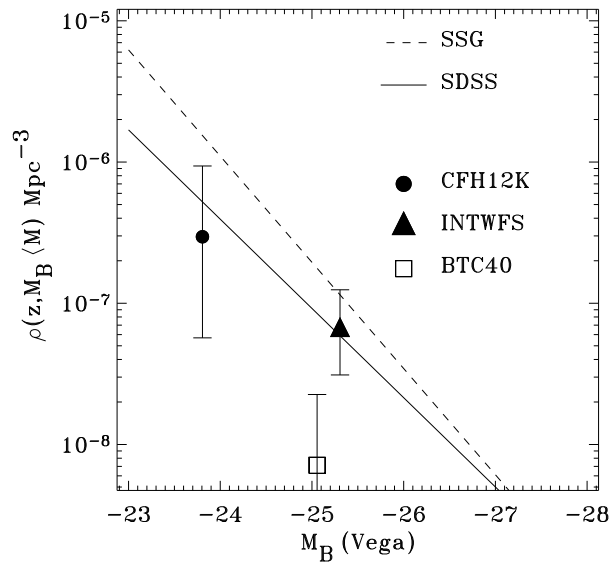,width=8cm}
 \caption{\label{cfht density} The quasar space density, inferred from
 three contemporary low luminosity high redshift surveys, is shown
 ($4.8<z<5.2$; CFH12K-this work, INTWFS-Sharp \etal (2001), Sharp
 (2002), BTC40-Monier \etal (2002)).  Predictions from two popular
 forms of the luminosity function are also shown, estimated as a
 function of limiting absolute magnitude, M$_B$(Vega), and evaluated
 at redshifts $z=5$ (SSG-Schmidt, Schneider and Gunn (1995), SDSS-Fan
 \etal (2001a)).  Error bars indicate the 65\% confidence intervals
 for each survey data set assuming Poisson statistics. Survey
 completeness corrections are applied to the CFH12K and INTWFS data
 points.  Survey limiting magnitudes correspond to CFH12K m$_z$=22.5,
 INTWFS m$_z$=21.0, BTC40 m$_i$=21.5.}
\end{figure}

\subsection{Improved selection criteria}
The plethora of early M star interlopers in candidate lists has dogged
quasars surveys at $z>4$.  Fundamentally, the spectral energy
distribution of early type M stars leads to colour indicies which
mimics those of quasars in the redshift range $4.8<z<5.8$.

When the spectroscopic stage of the survey is extended to include the
4.25\sqdeg\ NOAO field, the spectroscopic observations, reported in
section \ref{Observations : spectroscopic}, and the predicted colours
for low mass stars (Figure \ref{low mass stars iz}) argue in favour of
tighter colour constraints than are reported in section \ref{colour
selection}.  This would reduce the number of low mass early M stars
included in the candidate list at the expense of survey completeness.

In principle, extending the wavelength baseline of the colour
selection criteria allows a greater degree of separation between the
two source populations.  Fan \etal (2001b) perform targeted IR
observations of $i$ band drop out objects from SDSS (with repeated $z$
photometry to confirm the validity of the single optical band
photometry) to detect the first quasar at z$>$6.0. At such high
redshifts the stellar contaminants are L and T dwarfs with
intrinsically redder $I-Z$ and $Z-J$ colour indicies than M stars.  At
the lower redshifts probed by our CFH12K survey $4.8<z<5.2$ the
$z-J\sim0.5$ colour index of quasars is close to that of the early M
star contaminant ($0.7<z-J<1.3$ for M0V-M5V, Figure \ref{low mass
stars iz}).  With such a colour separation between the two source
populations, high accuracy photometry, often difficult to maintain
over wide fields of view at low luminosities, will be required if the
IR colour index is to be used in selection criteria.

Nevertheless, with the advent of large format IR mosaic cameras at
4meter class telescope facilities such as WIRCam at CFHT, WFCAM on
UKIRT (with the UKIDSS Large Area Survey (LAS) of particular
relevance\footnote{http://www.ukidss.org/}) and the VISTA survey
projects it will soon be possible routinely acquire wide field IR
photometry of sufficient depth and accuracy to introduce an additional
colour index to reduce the contamination by low mass stars.

\section{Conclusions}
In conclusion, we detect only a single high redshift, $4.8<z<5.2$,
quasar in the 1.8\sqdeg\ of the CFH12K survey currently investigated
spectroscopically.  The inferred space density of
2.96$\times$10$^{-7}$ Mpc$^{-3}$ (65\% confidence limits
$\pm_{7.5\times10^{-9}}^{1.6\times10^{-6}}$) is lower than that
predicted by the extrapolation of the Fan \etal (2001a) quasar
luminosity function (derived for m$_i$$<$19.6(Vega)) to fainter
magnitudes. 

The dearth of high redshift quasars in our 1.8\sqdeg\ survey area and
the quasar density reported by Sharp \etal (2001), found to be in
agreement with the luminosity of Fan \etal (2001a), is indicative of a
possible turn over in the luminosity function at faint quasar
magnitudes, although the detection of only a single redshift
$4.8<z<5.2$ quasar is consistent with the luminosity function at the
65\% confidence level.  Further evideance for such a turnover is
provided by Richards \etal (2003).  Richards \etal use the apparant
lack of multiply imaged quasars amongst a sample of four $z>5.7$
quasars, identified within SDSS, to constrain the bright end slope and
break point, M$_{B}^{*}$, of the luminosity function.  Their analysis
suggest a break point at M$_{B}^*=-24.38$(Vega)
(M$_{\rm{1450}}^*=-24.0$AB).  When $I$ band data from the public NOAO
survey become available, it will be possible to triple the areal
coverage of our CFH12K quasar survey.

It is worthy of note that all candidate objects have been identified
with intrinsically unresolved (PSF limited) stellar sources.  Both the
INTWFS (Sharp \etal (2001)) and to a lesser degree the BTC40 (Monier
\etal (2002)) quasar survey programs suffered from contamination of the
candidate list by compact galaxies.  As quasar surveys reach fainter
limiting magnitudes, one progresses further down the galaxy luminosity
function and the number of compact foreground galaxies becomes large.
The large volume of colour space occupied by normal galaxies, which
are not tied to a well defined colour space locus as is the case with
stars, means spectroscopic follow up of targets can become
prohibitively expensive unless a reliable star/galaxy classification
can be adopted.  The high spatial resolution and excellent image
quality of the CFH12K imaging system has resulted in the removal of
such extended sources from our candidate lists.

\section*{Acknowledgements}
RGS acknowledges the receipt of a PPARC studentship. Thanks are due to
Mike Irwin, for help and guidance during mosaic data reduction and the
construction of a world coordinate system, and Simon Hodgkin for
useful discussions regarding the spectral features of low mass stars.

Spectroscopic observations where obtained at the Gemini Observatory,
which is operated by the Association of Universities for Research in
Astronomy, Inc., under a cooperative agreement with the NSF on behalf
of the Gemini partnership: the National Science Foundation (United
States), the Particle Physics and Astronomy Research Council (United
Kingdom), the National Research Council (Canada), CONICYT (Chile), the
Australian Research Council (Australia), CNPq (Brazil) and CONICET
(Argentina)

\label{lastpage}

\end{document}